\documentclass[prb,preprint]{revtex4-1} 

\usepackage{amsmath}  
\usepackage{amsfonts} 
\usepackage{graphicx} 

\begin{document}


\title{Reconciling a Reactionless Propulsive Drive with the First Law of Thermodynamics}

\author{Andrew J. Higgins}
\email{andrew.higgins@mcgill.ca} 
\altaffiliation[permanent address: ]{817 Sherbrooke St. W., 
  Montreal, Quebec, Canada} 
\affiliation{Department of Mechanical Engineering, McGill University, Montreal, Quebec, Canada H3A-0C3}


\date{\today}

\begin{abstract}
A ``space drive'' is a hypothetical device that generates a propulsive force in free space using an input of power without the need for a reaction mass.  Any device that generates photons (e.g., a laser) would qualify as a propellantless ``photon rocket,'' but the force generated by emitting photons per power input (3.33 $\mu$N/kW) is too small to be a practical propulsion device.  The ability to generate greater force per power input would be highly desirable, but, as demonstrated in this paper, such a device would be able to operate as a perpetual motion machine of the first kind.  Since applying a constant force results in a constant acceleration, the kinetic energy of a mass driven by such a device increases quadratically with time, while the energy input increases only linearly with time.  Thus, at some point, the kinetic energy of the device-driven mass exceeds the energy input, and if this energy is collected via decelerating the mass (via regenerative electromagnetic braking, for example), then there would be a net gain in energy.  For devices with thrust-to-power ratios on the order of 1 N/kW that have been discussed recently in connection with the so-called EM drive, this breakeven occurs at velocities low enough to be feasible with current technology, clearly demonstrating the absurdity of such a device. When relativistic effects are taken into account, it is shown that the photon rocket can only reach energy breakeven as the accelerated mass asymptotically approaches the speed of light.  Thus, any device with a thrust-to-power ratio greater than the photon rocket would be able to operate as a perpetual motion machine of the first kind, and thus should be excluded by the First Law of Thermodynamics.
\end{abstract}

\maketitle 

\section{Introduction} 

A device that can generate thrust without consuming reaction mass (i.e., propellant) has long been sought by propulsion engineers as the so-called ``space drive.''  Such a device might be discounted as violating the conservation of momentum, however, this fact has not deterred a number of researchers from attempting to hypothesize and, in some cases, build such a device.  A number of recent reports of experimental investigation into thrust generation by RF resonant cavities have appeared accompanied with statements that suggest such devices may be able to generate 0.4 N of thrust per kW of electrical power input or greater.\cite{{Juan},{Brady}}  A number of elaborate explanations have been advanced, ranging from classical electromagnetics to interactions with virtual particles in the quantum vacuum, in order to explain this result and reconcile it with the conservation of momentum.  This paper will not address these issues whatsoever, but rather will examine the consequences in relation to the First Law of Thermodynamics if such a device were to exist.

The photon rocket is one example of a propellantless, thrust-generating device, and a great many people already own one in the form of a laser pointer.  Since photons possess momentum $p$ = $\frac{e}{c}$ where $e$ = $h \nu$, a device emitting photons along an axis will result in thrust being applied in the opposite direction.  The rate of change of momentum of the photon-emitting device if operating in free space, and thus the thrust $F$, is determined by the power input $P_{\mathrm{in}}$
\begin{equation}
\label{PhotonThrust}
F = \left( \text{momentum per photon} \right) \times \left( \text{rate of photon generation} \right) = \left( \frac{h \nu}{c} \right) \left( \frac{P_\mathrm{in}}{h \nu} \right)	= \frac{P_{\mathrm{in}}}{c}
\end{equation}
so the conversion factor between power and thrust, which will be denoted $\eta$, is
\begin{equation}
\label{PhotonNu}
\eta_{\mathrm{pr}} = \frac{F}{P_\mathrm{in}}=\frac{1}{c} = 3.336 \frac{\mu\mathrm{N}}{\mathrm{kW}}
\end{equation}
where the ``pr'' subscript refers to ``photon rocket.''  A typical laser pointer left on in free space would have a $\Delta V$ on the order of 0.1 mm/s by the time its battery ran out.

\section{Classical Analysis}

The recent claims regarding the EM drive are that the value of $\eta$ is much greater (i.e., five orders of magnitude greater) than those of the photon rocket, with values as great as $\eta_\mathrm{EM} = 0.4 \frac{\mathrm{N}}{\mathrm{kW}}$.\cite{{Juan},{Brady}}  This raises a troubling issue, however, in that the kinetic energy of a mass with a constant force applied will increase with the square of time, while the electrical energy input only increases linearly with time.  A mass with such a drive attached would eventually have a kinetic energy greater than the energy input.  It is possible to solve for when this occurs as follows.

The kinetic energy of a mass undergoing constant acceleration starting from rest is
\begin{equation}
\label{KE}
\text{\textit{KE}}  = \frac{1}{2} m V^2 = \frac{1}{2} m \left( \frac{F}{m} t \right)^2 = \frac{1}{2} m \left( \frac{\eta P_\mathrm{in}}{m} t \right)^2
\end{equation}
The energy input to the hypothetical drive, required to generate constant force, is
\begin{equation}
\label{Ein}
E_\mathrm{in} = P_{\mathrm{in}} t
\end{equation}
Equating these two expressions, we can solve for the time required for the kinetic energy to equal the energy input
\begin{equation}
\label{BEtime}
t_\mathrm{be} = \frac{2 m}{\eta^2 P}
\end{equation}
where ``be'' denotes ``breakeven.''  The velocity at this time is
\begin{equation}
\label{BEvelocity}
V_\mathrm{be} = \frac{2}{\eta}
\end{equation}
For a value of $\eta_\mathrm{EM} = 0.4 \frac{\mathrm{N}}{\mathrm{kW}}$, this breakeven velocity occurs at 5000 m/s.  Note that this velocity (which is somewhat less than orbital velocity) is sufficiently low that relativistic considerations are negligible.  Also note that this velocity is not beyond the theoretical capability of electromagnetic linear motors, and this allows us to entertain the notion of actually building such a machine.  For example, an evacuated track 10 km long and supplied with 1 GW of power that is fed via induction along the track to a 100 kg mass equipped with an EM drive would result in the mass experiencing an acceleration of 4000 $\mathrm{m/s}^2$ (i.e., 407 g's) and having a velocity of 8.94 km/s at the end of the track, at which point the mass would have a kinetic energy of 4 GJ.  The time to accelerate the mass is only 2.23 seconds, however, so the energy input required is only 2.23 GJ.  At the end of the track, the mass could be decelerated via regenerative braking, generating more energy out (4 GJ) than was input (2.23 GJ), for a net gain of 1.76 GJ free energy.

While it could be argued that inefficiencies in supplying power to the EM drive or in extracting power via electromagnetic braking would result in a net loss, these considerations only represent engineering challenges.  Such considerations do not apply when constructing a perpetual motion machine as a conceptual device in order to prove that it violates the First Law. (Note that, historically, the separation of losses into those that could be avoided with more elegant engineering, such as friction and electrical resistance, and those that are unavoidable, such as a heat engine needing to reject heat to a cooler reservoir, was the key insight by Sadi Carnot that enable the development of thermodynamics as a science.\cite{Baeyer})  Further, note that the result Eq.~(\ref{BEvelocity}) is the breakeven velocity; the net gain of the device can be increased arbitrarily by allowing the mass to reach greater velocities via a longer acceleration track and offset any component inefficiencies.

\section{Relativistic Analysis}

It is interesting to return to the traditional photon rocket, with $\eta_\mathrm{pr} = \frac{1}{c}$, and examine the resulting breakeven velocity.  We find that that Eq.~(\ref{BEvelocity}) predicts that a photon rocket, fed via an external power source, would reach a velocity where the kinetic energy exceeds the energy input at $V_\mathrm{be} = 2 c$.  Clearly, this result necessitates a relativistic analysis for the photon rocket.

The analysis will parallel that in Section II.  If an object with a rest mass $m$ experiences a constant force, then 
\begin{equation}
\label{Fdpdt}
F = \frac{\mathrm{d}p}{\mathrm{d}t}
\end{equation}
where the relativistic momentum is given by
\begin{equation}
\label{momentum}
p = \frac{m V}{\sqrt{1-\frac{V^2}{c^2}}}
\end{equation}
Thus, the acceleration is
\begin{equation}
\label{acceleration}
a = \frac{\mathrm{d} V}{\mathrm{d} t} = \frac{F}{m} \left( 1 - \frac{V^2}{c^2} \right )^\frac{3}{2}
\end{equation}
and the velocity as a function can be obtained by solving this differential equation as
\begin{equation}
\label{integrateacceleration}
V(t) =\left ( \frac{1}{\sqrt{1+\frac{(\frac{F}{m})^2 t^2}{c^2}}} \right) \frac{F}{m} t
\end{equation}
This equation can be inverted to find the time required to reach a velocity
\begin{equation}
\label{timeforvelocity}
t(V) = \left(\frac{1}{\sqrt{1-\frac{V^2}{c^2}}} \right) \frac{V}{\frac{F}{m}}
\end{equation}
The energy input to the device is still given by the relation (cf. Eq.~(\ref{Ein}))
\begin{equation}
\label{EinRel}
E_\mathrm{in} = P_{\mathrm{in}} t(V) = \left(\frac{1}{\sqrt{1-\frac{V^2}{c^2}}} \right) \frac{V}{\frac{\eta}{m}}
\end{equation}
where we have used $F = \eta P_\mathrm{in}$.  For the case of the photon rocket, $\eta_\mathrm{pr}=\frac{1}{c}$, the energy input require to reach a speed $V$ is
\begin{equation}
\label{EinPhotonRocket}
E_{\mathrm{in}_\mathrm{pr}} = \left(\frac{\frac{V}{c}}{\sqrt{1-\frac{V^2}{c^2}}} \right) m c^2
\end{equation}
The relativistic kinetic energy is given by
\begin{equation}
\label{KERel}
\text{\textit{KE}} = \frac{m c^2}{\sqrt{1-\frac{V^2}{c^2}}} - m c^2
\end{equation}

We can plot these two energies as a function of the final velocity, as is done in Fig. 1.  The energy input is always greater than the kinetic energy of the device for all subluminal speeds.  The ratio of the energies in the limit as the velocity approaches the speed of light is
\begin{equation}
\label{LimitLightspeed}
\lim_{V \to c} \frac{E_{\mathrm{in}_\mathrm{pr}}}{\text{\textit{KE}}_\mathrm{pr}} = 1
\end{equation}
Thus, the kinetic energy of photon rocket approaches the energy input only as its velocity approaches that of light. This result means that it would not be worthwhile to build a device like that discussed in Section II using a photon rocket as a means to generate net energy, even if we had the technology to feed power to the photon rocket as it approaches arbitrarily close to the speed of light.

\begin{figure}[h!]
\centering
\includegraphics{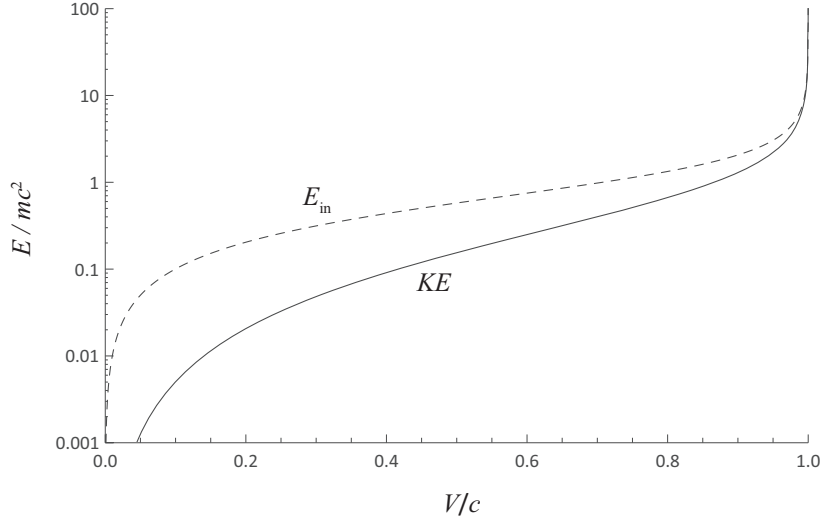}
\caption{Comparison of the kinetic energy of a photon rocket as a function of velocity (normalized by the speed of light) compared to the energy input required to reach that velocity.  Energy is normalized by the rest mass of rocket and the speed of light.}
\label{energycomparison}
\end{figure}

Equating Eq.~(\ref{EinRel}) and  Eq.~(\ref{KERel}), the relativistic breakeven velocity can be solved for devices other than photon rockets
\begin{equation}
\label{RelBreakeven}
\frac{V_{\mathrm{be}}}{c} = \frac{2 \frac{\eta}{\eta_\mathrm{pr}}}{1 + \left ( \frac{\eta}{\eta_\mathrm{pr}} \right)^2}
\end{equation}
where the ratio $\frac{\eta}{\eta_\mathrm{pr}}$ represents the device's thrust-to-power ratio normalized by the value of the photon rocket.  It can be seen that, for any value of $\eta > \eta_\mathrm{pr}$ = 1/$c$, the breakeven velocity occurs at less than light speed.  Thus, we are left with the conclusion that \emph{any} device with a value of thrust-to-power greater than that of the photon rocket is, in principle, capable of being operated as a perpetual motion machine of the first kind at velocities less than light speed.

\section{Conclusions}

The fact that the EM drive, or any other reactionless drive that has a thrust-to-power ratio greater than a photon-emitting device, would enable a perpetual motion machine of the first kind suggests that such a device cannot exist.  This objection is not as easily explained away as the conservation of momentum objection to a reactionless drive, because this result suggests than a source of free and infinite energy is already at our technological disposal.  Any conditions placed on the operation of the hypothetical ``space drive'' in order to make is consistent with the First Law would also render it useless as a propulsion device; if it can work as a propulsion device, it can also function as a perpetual motion machine of the first kind.  Further investment into investigating this concept should be tempered by the history of attempts to realize perpetual motion machines.


\begin{thebibliography}{99}

\bibitem{Juan}  Yang Juan, Wang Yu-Quan, Ma Yan-Jie, Li Peng-Fei, Yang Le, Wang Yang, and He Guo-Qiang,
``Prediction and experimental measurement of the electromagnetic thrust generated by a microwave
thruster system,'' Chinese Physics B, \textbf{22} (5), 050301 (2013).

\bibitem{Brady} David A. Brady, Harold G. White, Paul March, James T. Lawrence, and Frank J. Davies,
``Anomalous Thrust Production from an RF Test Device Measured on a Low Thrust Torsion Pendulum,''
50th AIAA/ASME/SAE/ASEE Joint Propulsion Conference, 28--30 July 2014, AIAA 2014-402 (2014).

\bibitem{Baeyer}  Hans Christian Von Baeyer, \textit{Warmth Disperses and Time Passes: The History of Heat} (Modern Library, 1999), p.~35-46.


\end{thebibliography}
\end{document}